# PROGRESS IN HIGH FIELD ACCELERATOR MAGNET DEVELOPMENT BY THE US LHC ACCELERATOR RESEARCH PROGRAM*

Gian Luca Sabbi (LBNL) for the LARP collaboration

*Abstract*

The maximum magnetic field available to guide and focus the proton beams will be the most important factor driving the design of the High Energy LHC. The US LHC Accelerator Research Program (LARP) is a collaboration of US National Laboratories aiming at demonstrating the feasibility of $Nb_3Sn$ magnet technology for application to future colliders. While LARP is primarily focused on the requirements of the High-Luminosity LHC (HL-LHC), it is also directly relevant to the High-Energy LHC (HE-LHC). Program results and future directions will be discussed.

## INTRODUCTION

A series of upgrades to the LHC and its injectors is under study to achieve a significant increase of the luminosity with respect to the baseline design [1]. Replacing the first-generation IR quadrupoles with higher performance magnets is one of the required steps in this direction. Although designs based on NbTi conductor are being considered, the intrinsic properties of $Nb_3Sn$ make it a strong candidate to meet the ultimate performance goals in terms of operating field, temperature margin, and radiation lifetime. Under typical upgrade scenarios, the new magnets will provide increased focusing power to double or triple the luminosity, and at the same time will be able to operate under radiation loads corresponding to a 10-fold increase in peak luminosity, and with radiation lifetime consistent with a 3000 $fb^{-1}$ integrated luminosity goal.

Starting in 2004, the LHC Accelerator Research Program (LARP) collaboration has led the US effort to develop $Nb_3Sn$ quadrupole magnets for the LHC luminosity upgrade [2]. The program is founded on the knowledge base and infrastructure of the DOE General Accelerator Development programs at BNL, FNAL and LBNL. With respect to these programs, it provides specific focus and resources to select the best available technologies for the luminosity upgrade and bridge the gap from proof-of-principle models to fully developed prototypes incorporating all features required for operation in the LHC accelerator. Significant progress has been made to date and the program is well positioned to complete the technology demonstration by 2014 and initiate a construction project. A successful luminosity upgrade based on $Nb_3Sn$ will represent a significant step toward a High-Energy LHC and open the way to other applications within and outside high energy physics.

*Work supported by the US Department of Energy

## HIGH FIELD MAGNET TECHNOLOGIES

Excellent mechanical and electrical properties of multi filamentary NbTi have made it the conductor of choice in all superconducting accelerators to date. However, the intrinsic properties of NbTi limit its field reach in practical accelerator applications to about 8 T. In order to surpass this threshold, superconductors with higher upper critical field are needed. Niobium-Tin ($Nb_3Sn$) is currently the most advanced material for practical applications. It carries current densities similar to NbTi at more than twice the field, and is available in long lengths with uniform properties. $Nb_3Al$ offers lower strain sensitivity with respect to $Nb_3Sn$, but its manufacturing process is not sufficiently well developed to support magnet fabrication. The low-temperature properties of HTS materials such as Bi-2212 are far superior to both $Nb_3Sn$ and $Nb_3Al$. However, many technology challenges need to be addressed before practical designs can be developed and implemented in prototypes.

All superconductors suitable for high field applications are brittle and strain sensitive, requiring new approaches to magnet design and fabrication to complement or replace those established for NbTi. In particular, because of their brittleness, high field superconductors cannot be drawn to thin filaments like NbTi, but have to be formed in the final geometry by high-temperature heat treatment. In the fully reacted state, the filaments are extremely sensitive to strain. Therefore, attempting to wind pre-reacted cables in accelerator-type coils would result in unacceptable critical current degradation at the ends. Instead, coils are wound using un-reacted cable, when components are still ductile, and the superconductor is formed by high temperature heat treatment after coil winding. This technique requires the use of insulation and coil structural components that can withstand the high reaction temperatures. In addition, new approaches to mechanical support and quench protection are required to safely handle reacted coils through magnet assembly, cool down and excitation

A significant and sustained R&D effort is required to develop technologies that can take advantage of the properties of high field superconductors while coping with the associated challenges. Early work on $Nb_3Sn$ accelerator magnets was performed at BNL [3], CEA [4], CERN [5-6], and LBNL [7]. In the mid-90s, the dipoles MSUT (Twente University) and D20 (LBNL) reached fields of 11-13 T [8-9]. More recently, the LBNL dipoles RD3-B and HD1b achieved record field of 14.7 T and 16.1 T, respectively, using simple racetrack coil designs

[10-11]. The LARP program was established to build on this base and develop the technology to a mature state, consistent with the requirements of the High-Luminosity LHC project.

## THE LARP PROGRAM

### Goals and organization

LARP was established in 2004 to enable active participation of the U.S. scientific community in the accelerator research program of the LHC machine. While the program scope included accelerator commissioning and operation, special emphasis was given to the development of magnet technologies relevant to the LHC luminosity upgrade, which was recognized as one of the highest physics priorities by the US HEP advisory panel [12]. LARP is also intended to serve as a vehicle to advance collaboration among US Laboratories as well as international cooperation in large science projects.

The documents that initiated the program identified its key goals, to be achieved in close collaboration with CERN:

- Help the LHC achieve its design luminosity quickly, safely and efficiently.
- Continue to improve LHC performance by advances in understanding and the development of new instrumentation.
- Use the LHC effectively as a tool to gain a deeper knowledge of accelerator science and technology.
- Extend LHC as a frontier High Energy Physics instrument with a timely luminosity upgrade.

LARP was firmly established as an advanced R&D program, which would help the US HEP community in maintaining a leadership role in accelerator technology, and set the basis for a separately funded construction project. "Preparing to build the next generation hadron collider" was also explicitly mentioned among the key program goals in the LARP proposal (Fig.1).

The program is organized in three sections: (i) accelerator systems, (ii) magnet systems and (iii) programmatic activities. The accelerator systems section includes the development of advanced instrumentation and collimation systems, as well as accelerator physics studies. The magnet systems section is focused on the development of $Nb_3Sn$ interaction region quadrupoles, and is described in detail in this paper. The programmatic activities section manages the long term visitor program and the Toohig post-doctoral fellowship.

| Deliverables \ Goals | Hardware Commissioning | Beam Commissioning | Fundamental Accelerator Research | Instrumentation & Diagnostics | Magnet R&D |
|---|---|---|---|---|---|
| Maximize HEP at the LHC | Y | Y | Y | Y | |
| Improve LHC Performance | | | Y | Y | |
| Advance Accelerator Science & Technology | | | Y | Y | Y |
| Extend LHC HEP by a Timely Upgrade | | | Y | Y | Y |
| Prepare to Build the Next Generation Hadron Collider | Y | Y | Y | Y | Y |

Fig. 1: LARP goals and deliverables matrix [2]

### Magnet program components

The LARP magnet program was conceived as a progression of studies and technological steps, starting from simple systems designed to address specific R&D issues, and building toward more complex configurations incorporating all required features for operation in the accelerator. The program organization reflects this approach and has evolved in time to adapt to the different stages of the R&D. The main areas, corresponding to "level 2" categories in the work breakdown structure, are:

- Materials R&D, including: strand specifications, procurement and characterization; cable fabrication, insulation and qualification; coil heat treatment optimization and verification.
- Technology development with racetrack coils. This area was a key component of the program from its inception until 2008. Through the Sub-scale Quadrupole (SQ) and Long Racetrack (LR) models, it addressed fundamental issues of conductor performance, mechanical analysis, instrumentation quench protection, and most notably, scale-up of coil and structures to 4 m length, paving the way to the long quadrupole program.
- Design studies: This area was also very active in the first part of the program, to select the most promising designs for future model quadrupoles, compare different IR layouts, and perform supporting studies in areas such as radiation deposition and field quality. While the program has progressively shifted toward experimental demonstrations, renewed focus on this area is developing in connection with the HL-LHC design study [13].
- Model quadrupoles: this area oversees the detailed design, fabrication and testing of short quadrupole models, including the 90 mm aperture Technology Quadrupoles (TQC and TQS) and the 120 mm aperture High Field Quadrupoles (HQ).
- Long Quadrupoles (LQ), which covers the scale up from 1 m to 4 m length (LQ and LHQ models).

Each area is organized around tasks with specific goals and milestones. Individual task typically utilize expertise, resources and infrastructure from several laboratories, leading to close collaboration at the level of each activity. This approach may appear less efficient with respect to a project-type organization in which responsibilities for key deliverables are distributed among laboratories, with each group working independently on its portion. However, it has proven extremely valuable in comparing and integrating the experience and methods developed by different groups, and a key element of the program success both from a technical and collaboration standpoint.

*Fabrication and test database*

Since its start in 2004, the LARP program has fabricated and tested a large number of models of different designs. This section summarizes the tests performed and the key issues addressed. Progress and issues in each area are summarized in the following section.

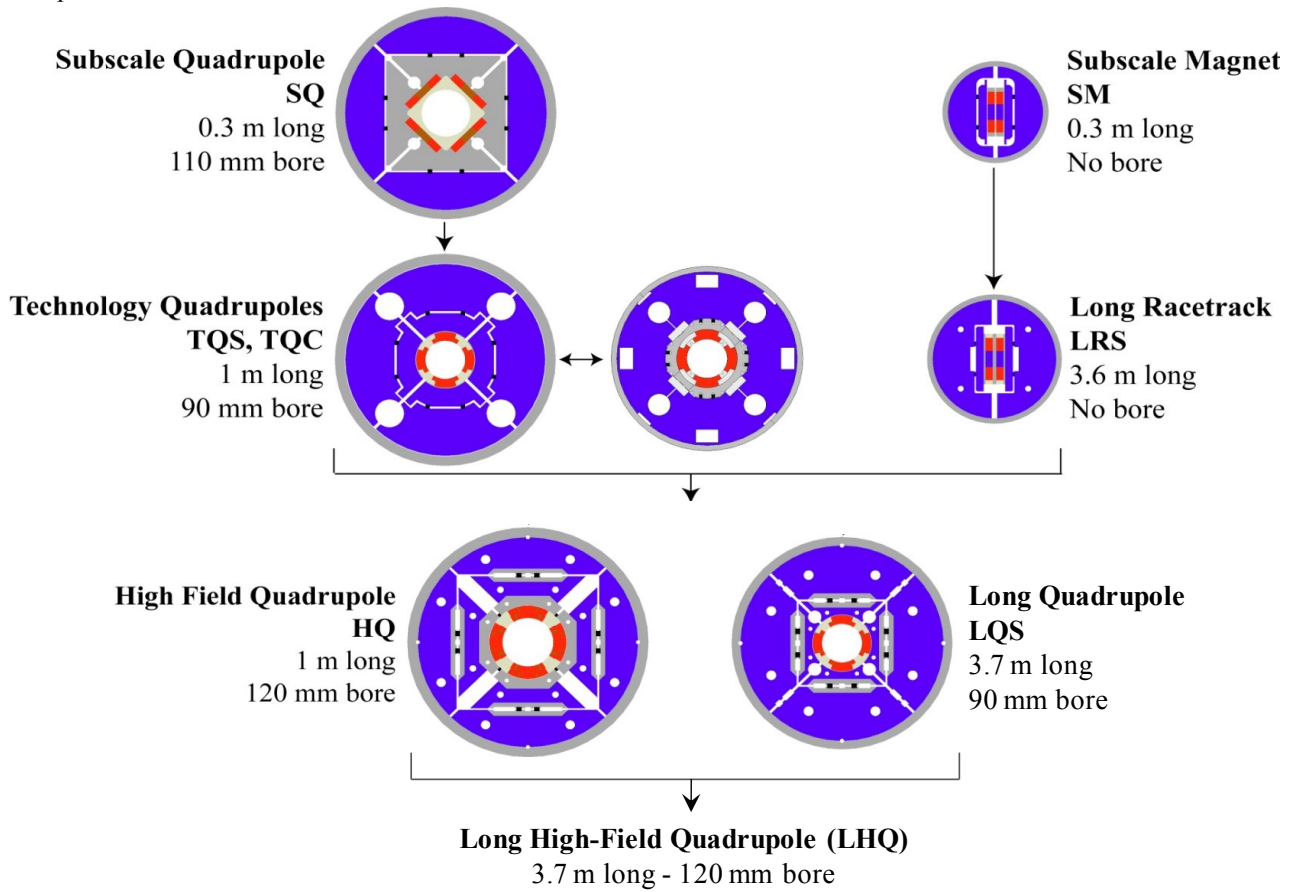

Fig. 2: LARP magnet development flow-chart

Figure 2 is a magnet development flowchart showing the LARP model magnets and their progression from technological tests toward accelerator quality designs. The main program components are:

1. Sub-scale Quadrupole - SQ (LBNL, FNAL). SQ is based on four racetrack coils of the LBNL "sub-scale" design [14]. A combination of existing and new coils was used leading to five tests at 4.5 K and two tests at 1.9 K [15-16]. Among the highlights of these tests were:
   - Demonstration of conductor performance up to the short sample limits under conditions similar to those required by the Technology Quadrupoles (field, current, stress) and using the same heat treatment.
   - Detailed 3D finite element modeling and verification of stress calculation against strain gauge measurements.
   - Studies of quench propagation and protection, including temperature and stress limits during a quench.
   - Studies of the effect of axial pre-load on the quench performance and training.

   In addition, the SQ tests indicated that block-coil quadrupoles can perform at the expected levels in practical configurations. However, cos(2θ) coils were selected for the LHC IR quad application since design studies showed that they would provide significantly better magnetic efficiency for this application.

2. Sub-scale Magnet - SM (BNL, LBNL). This magnet was used as a technology transfer tool in preparation

for the design and fabrication of the long racetrack coils at BNL. Two sub-scale coils were fabricated and assembled at BNL using design, cables, parts, mechanical structure and fabrication procedures provided by LBNL. The magnet was also tested at BNL and achieved its full conductor potential [17].

3. Long Racetrack Shell - LRS (BNL, LBNL). The main goal of LRS was to provide a first demonstration that the $Nb_3Sn$ coils and shell based structures could be scaled to lengths significantly higher than 1 meter. The coil design was very similar to the sub-scale magnet, with a length increase of more than a factor of ten. The support structure was designed and pre-assembled at LBNL. Two coils were fabricated, assembled and tested at BNL achieving 91% of the short sample limit [17]. Based on feedback from this test, the support structure which originally utilized a one-piece shell was subdivided in four sections, leading to further performance improvements (96% of SSL) in the second test using the same coils [18].

4. Technology Quadrupole – TQ (FNAL, LBNL + CERN). The TQ models are based on the traditional $\cos(2\theta)$ coil design with 90 mm aperture and 1 m length. Three generations of coils were fabricated using different wire designs. In total, more than 30 coils were fabricated using a distributed production line, with winding/curing performed at FNAL and reaction/impregnation performed at LBNL. Two support structures were compared, a collar-based structure designed by FNAL and a shell based structure designed by LBNL. About 15 models were tested in a variety of configurations at LBNL, FNAL and CERN [19-20]. Among the main studies and results obtained using the TQ models are:
   - Achieved 240 T/m in 90 mm aperture, about 20% higher than the performance target.
   - Demonstrated robust performance, in particular the capability to transport, disassemble and reassemble coils in different configurations.
   - Performed a systematic investigation of $Nb_3Sn$ stress limits (engineering design space)
   - Completed a fatigue test involving 100 cycles from low to high current.

5. Long Quadrupole Shell – LQS (BNL, FNAL, LBNL). LQS is a scale-up of the TQS design from 1 m to 4 m. The development of long $Nb_3Sn$ quadrupoles was recognized as a key R&D goal from the program outset. In April 2005, LARP, DOE and CERN agreed that achieving a gradient of 200 T/m in a 90 mm aperture, 4 m long quadrupole would serve as a convincing demonstration of such scale-up. The primary purpose of both TQ and LR programs was to serves as a basis for LQ. All three labs participated in the LQ design, fabrication and test activities. The 200 T/m target was achieved during the first test in December 2009 [21]. A second test with optimized preload using the same coils (LQS01b) achieved a 10% increase in performance, to 220 T/m. The next step is the assembly and test of LQS02, using four new coils, to demonstrate reproducibility. A third series of tests is also planned using the latest generation conductor (RRP 108/127).

6. High-Field Quadrupole - HQ (BNL, FNAL, LBNL + CERN). Detailed optics and layout studies of the upgraded LHC insertions indicate that increasing the quadrupole aperture leads to improved performance. Taking into account the space limitations in the tunnel, an aperture of 120 mm was selected for the development of upgraded quadrupole models based on NbTi. In order to explore the technological limits associated with larger aperture, and to provide a direct comparison between NbTi and $Nb_3Sn$ performance, the same aperture was selected by LARP for the next series of High-Field Quadrupoles. The 120 mm aperture, two-layer coil design using a 15 mm wide cable results in a 15 T peak field and 1.2 MJ/m stored energy, about a factor of 3 higher than in TQ and LQ. For the first time in LARP, coil alignment features are included at all phases of fabrication, assembly and excitation. To date, 12 coils have been fabricated and 3 tests were performed. During the first test [22] the magnet achieved 155 T/m at 4.5 K, well above the intrinsic limit of NbTi at 1.9 K. However, high rates of insulation failures were observed, prompting a revision of the cable and coil design to decrease stress during fabrication. A scale up of the HQ design to 4 m length is planned as a final technology demonstrator.

## R&D PROGRESS AND ISSUES

### Strand design and fabrication

Three wire types were utilized in LARP, all produced by Oxford Superconducting Technology (OST):
- Modified Jelly Roll wire with 61 sub-elements, 54 of which contain superconducting filaments while the remaining 7 are made of copper stabilizer (MJR 54/61)
- Rod Restack Processed wire with 61 sub-elements, 54 of which contain superconducting filaments while the remaining 7 are made of copper stabilizer (RRP 54/61)
- Rod Restack Processed wire with 127 sub-elements, 108 of which contain superconducting filaments while the remaining 19 are made of copper stabilizer (RRP 108/127)

The MJR wire represents an older generation wire that was already retired from production at the beginning of the program. It was used in the first generation TQ models since it was available in sufficient quantity to allow a direct comparison of different mechanical structures.

The RRP 54/61 was used in the majority of the LARP tests to date. It delivered solid performance allowing the

LR, TQ and LQ models to reach their R&D objectives and performance goals. However, this design results in a rather large effective filament size (~70 μm) in the strand diameter of interest (0.7-0.8 mm) leading to stability thresholds which are only within a factor of 2 above the operating point. Further erosion of the stability margin may result from conductor degradation due to processing or strain. As a result, performance limitations have been observed for moderate field designs at low temperature.

The RRP 108/127 was first procured by LARP in 2007, when it was still considered an R&D wire by OST, to evaluate its performance and encourage further development and transition to the production stage. It provided solid performance in the TQS03 model with no signs of instability, leading to its adoption as a baseline LARP wire starting in 2009, However, due to the long lead times for procurement and magnet fabrication, the first models to benefit from this transition will only be tested in 2012. In addition, further improvements to the 108/127 design are required to match the average piece length and critical current densities obtained in the 54/61 design. The 5-6 year cycle from initial evaluation to full utilization in the magnet fabrication pipeline indicates that incorporating newer generations of wire (such as RRP 217 or Powder-in-Tube) before the 2015 anticipated start of IR quadrupole production will be a challenge.

### Cable design and fabrication

Although the fabrication of Nb3Sn cables was already well established at the start of the program, LARP provided an opportunity for larger scale manufacturing, optimization and characterization. To date, more than 7 km of cable of three different designs were fabricated with minimal losses. The current R&D effort is focusing on transitioning from a three-step process involving a first cable fabrication pass at larger size, followed by anneal and re-roll to final size, to a one-step process using pre-annealed strand. The one-step process is expected to be more robust and efficient, and is compatible with the introduction of thin cores for control of the AC losses. Several cored cables have been fabricated for the latest generation HQ models using stainless steel and fiberglass cores. Coils have been fabricated using cored cables and will be assembled and tested in the near future.

### Quench performance and training

The capability to approach the full conductor performance in model magnets is an important indicator of the maturity of the technology, and the capability to reach the design point with minimal training and no retraining is essential requirement for operation in the accelerator. On both fronts, positive results were obtained. The full conductor potential, based on critical current measurements of extracted strands, without factoring in stress degradation, has been obtained in the best SQ, LR, TQ and LQ models at 4.5 K, indicating that the design and fabrication process is well understood and optimized. The best models also showed fast training and no retraining. However, new designs tend to require several iterations in order to achieve the best results. The steady process of systematic analysis and improvement defines the success of an R&D program like LARP, but it is clear that more work is needed to achieve full control of this technology, in particular for what concerns the coil design and fabrication, and especially the reaction step.

### Mechanical design and stress limits

Providing adequate mechanical support in high-field magnets based on brittle superconductors requires structures that can generate large forces while minimizing stress on the conductor at all stages of magnet fabrication and operation. Consistent with the R&D goals of the program, the application of new concepts and advanced modeling capabilities was emphasized. In particular, support structure originally developed at LBNL for high field dipoles [23] was applied to the LARP quadrupoles. This concept is based a thick aluminum shell, pre-tensioned a room temperature using water-pressurized bladders and interference keys. During cool-down, the stress in the shell increases due to differential thermal contraction relative to the iron yoke. This shell-based structure was evaluated against the more traditional collar-based structure in the TQ models, scaled-up to 4 m length in the LR and LQ models, and further optimized in the HQ models.

A series of tests were performed at CERN using TQS03 models to better understand the $Nb_3Sn$ stress limits and its tolerance to a large number of cycles [24]. It was found that the magnet could perform satisfactorily up to 200 MPa average coil stress, which results in peak local stresses of the order of 250 MPa. This result considerably expands the engineering design space with respect to what was previously considered as the limit of 150 MPa. In addition, a cycling test involving one thousand ramps from low to high field was performed, and no degradation was found.

### Alignment and Field Quality

Due to large beam sizes in the IR quadrupoles, their field quality plays a critical role on the beam dynamics during collision. Therefore, precise coil fabrication and structure alignment are required. Although early LARP magnets had limited alignment features, steady progress has been made and the last generation of HQ models incorporates full alignment at all steps of coil fabrication, magnet assembly and operation. No negative impact on mechanical support and quench performance resulting from the introduction of these features has been observed so far.

Field errors at injection are less critical, but need to be carefully analyzed since $Nb_3Sn$ wires exhibit large magnetization due to high critical current density and large filament size. Compensation of persistent current effects by saturation of carefully iron inserts may provide an intermediate solution. Ultimately, wires with larger number of sub-elements should be developed to decrease the effective filament size.

As previously mentioned, cored cables are being introduced in the HQ models to better control the distortions generated by eddy current during a ramp.

*Coil fabrication technology*

Several factors contributed to a steady improvement in coil fabrication procedures throughout the program. Different experiences and methods had to be compared and integrated in order to develop tooling and procedures that would be acceptable to all groups. Robust handling and shipping tools had to be devised to allow distributed coil production lines for the TQ, LQ and HQ models. Careful analysis was performed in relation to the scale up to 4 m length in the LR and LQ models. Nevertheless, a complete modeling framework is still not available, particularly in relation to the reaction process. The coil fabrication methods are still largely based on empirical knowledge and several iterations are typically needed to optimize new designs.

## LARP RELEVANCE TO HE-LHC

As previously noted, large portions of the magnet R&D effort performed by the LARP program in support of the LHC luminosity upgrade has direct relevance to the High-Energy LHC. Although the stated goal of 20 T field is beyond reach for $Nb_3Sn$, it is expected that a hybrid dipole design will be used with $Nb_3Sn$ providing a large portion of the total field. Among the key contributions of LARP to the development of technologies applicable to HE-LHC are:

- Scale up to long magnets: cable fabrication and QA, coil and structure fabrication, magnet assembly.
- Instrumentation and analysis
- Field quality, quench protection
- Accelerator integration: cooling, helium containment, alignment
- Development of radiation tolerant components
- Initial feedback on series production and operation issues:
- Infrastructure, production steps/times/cost
- Reliability, failure rates in production/operations

In addition, the capability to organize and integrate an effective R&D effort across Laboratories is a key contribution of LARP.

At the same time, it is clear that large portions of the R&D needed for HE-LHC are not covered by LARP. In particular, LARP is not involved with HTS technologies that will be required to push the field beyond 15 T. In addition, the small aperture required to limit magnet size and cost in HE-LHC will drive the magnet design in different directions with respect to those adopted for the large-aperture IR quadrupoles. Finally, an additional length scale-up of a factor of 2-3 will be required to achieve dipole lengths comparable to those used in the baseline LHC. Coupled with the small magnet aperture, this will also require incorporating a small sagitta in the magnet fabrication.

## SUMMARY

Intensive magnet R&D efforts are needed to meet the requirements of future colliders at the energy frontier. The LHC luminosity upgrade provides the opportunity to refine the results obtained in proof-of-principle $Nb_3Sn$ models and extend them to full-size production magnets suitable for operation in a challenging accelerator environment. The LARP program has made considerable progress in this direction, and is expected to complete the technology demonstration within the next several years. Successful construction and implementation in the high luminosity LHC will provide a stepping stone for the application of high field magnet technology to next generation colliders such as the High Energy LHC.